\documentclass[12pt]{article}
\usepackage{epsf}

\begin{document}
\setlength{\parskip}{2ex}
\setlength{\parindent}{0em}
\setlength{\baselineskip}{3ex}
\newcommand{\onefigure}[2]{\begin{figure}[htbp]
         \caption{\small #2\label{#1}(#1)}
         \end{figure}}
\newcommand{\onefigurenocap}[1]{\begin{figure}[h]
         \begin{center}\leavevmode\epsfbox{#1.eps}\end{center}
         \end{figure}}
\renewcommand{\onefigure}[2]{\begin{figure}[htbp]
         \begin{center}\leavevmode\epsfbox{#1.eps}\end{center}
         \caption{\small #2\label{#1}}
         \end{figure}}
\newcommand{\comment}[1]{}
\newcommand{\myref}[1]{(\ref{#1})}
\newcommand{\secref}[1]{sec.~\protect\ref{#1}}
\newcommand{\figref}[1]{Fig.~\protect\ref{#1}}
\newcommand{\mathbold}[1]{\mbox{\boldmath $\bf#1$}}
\newcommand{\mJ}{\mathbold{J}}
\newcommand{\momega}{\mathbold{\omega}}
\newcommand{\bz}{{\bf z}}
\def\bbbz{{\sf Z\!\!\!Z}}
\newcommand{\PP}{\mbox{I}\!\mbox{P}}
\def\sl2z{SL(2,\bbbz)}
\newcommand{\bbbq}{I\!\!Q}
\newcommand{\be}{\begin{equation}}
\newcommand{\ee}{\end{equation}}
\newcommand{\bea}{\begin{eqnarray}}
\newcommand{\eea}{\end{eqnarray}}
\newcommand{\nn}{\nonumber}
\newcommand{\unit}{1\!\!1}
\newcommand{\half}{\frac{1}{2}}
\newcommand{\shalf}{\mbox{$\half$}}
\newcommand{\transform}[1]{
   \stackrel{#1}{-\hspace{-1.2ex}-\hspace{-1.2ex}\longrightarrow}}
\newcommand{\inter}[2]{\null^{\#}(#1\cdot#2)}
\newcommand{\lprod}[2]{\vec{#1}\cdot\vec{#2}}
\newcommand{\mult}[1]{{\cal N}(#1)}
\newcommand{\Bn}{{\cal B}_N}
\newcommand{\B}{{\cal B}}
\newcommand{\Beight}{{\cal B}_8}
\newcommand{\Bnine}{{\cal B}_9}
\newcommand{\Eman}{\widehat{\cal E}_N}
\newcommand{\C}{{\cal C}}
\newcommand{\Q}{Q\!\!\!Q}
\newcommand{\comp}{C\!\!\!C}

\noindent

\thispagestyle{empty}
{\flushright{\small AEI-1999-24\\MIT-CTP-2896\\hep-th/9910054\\}}

\vspace{.3in}
\begin{center}\Large {\bf Del Pezzo Surfaces and Affine 7-brane Backgrounds} 
\end{center}

\vspace{.1in}
  \begin{center}
{\Large Tam\'{a}s Hauer}
\vspace{.05in}

{ {\it Max-Planck-Institut f\"ur Gravitationsphysik,\\
Albert-Einstein-Institut, D-14476 Golm, Germany}}
\vspace{.1in}

{\large and}

\vspace{.1in}

{\Large Amer Iqbal}

\vspace{.05in}

 { {\it Center for Theoretical Physics,\\
Department of Physics\\
MIT, Cambridge, Massachusetts 02139.\\}}
\vspace{.2in}

E-mail: {\tt hauer@aei-potsdam.mpg.de, amer@mit.edu}
\end{center}
\begin{center}October 1999\end{center}

\vspace{0.1in}

\begin{abstract}
A map between string junctions in the affine 7-brane backgrounds and
vector bundles on del Pezzo surfaces is constructed using mirror
symmetry.  It is shown that the lattice of string junctions with
support on an affine 7-brane configuration is isomorphic to the
K-theory group of the corresponding del Pezzo surface. This
isomorphism allows us to construct a map between the states of the
${\cal N}=2$, D=4 theories with $E_{N}$ global symmetry realized in
two different ways in Type IIB and Type IIA string theory. A subgroup
of the $\sl2z$ symmetry of the $\widehat{E}_{9}$ 7-brane background
appears as the Fourier-Mukai transform acting on the D-brane
configurations realizing vector bundles on elliptically fibered ${\cal
B}_{9}$.

\end{abstract}

\newpage

\section{Introduction}

The study of D-branes has provided important insights into
non-perturbative aspects of string theory as well as supersymmetric
gauge theories. Supersymmetric field theories are realized in string
theory either as the effective world-volume theory of a configuration
of D-branes or as the theory on the transverse space after
compactification on some appropriate manifold. In the later case
D-branes wrapped on various cycles of the compactification manifold
give rise to particles and non-critical strings in the gauge
theory. Thus D-branes provide a geometrical description of states in
the supersymmetric gauge theories.

Along the process of realizing QFT's as effective models in string
theory, previously unknown exotic SUSY theories have also been found.
An example is the ${\cal N}=1$ six dimensional $E_8$ theory
\cite{witten,GH,SW,witten-phase,MV-2,ganor,GMS} whose compactification  
on a torus leads to ${\cal N}=2$ theories with $E_N$ global symmetry
\cite{KMV2}. These four dimensional theories and their spectra are the
focus of our paper. Since there are different ways of obtaining
effective field theories in string theory, some models can be realized
in different ways. Our aim is to compare two different realizations of
the same theory, namely $d=4$, ${\cal N}=2$ SYM theory with $E_N$
global symmetry which arises both as the worldvolume theory of a
threebrane in IIB (or F-theory) \cite{vafa,sen,BDS,DKV} near a
particular set of 7-branes, and as the compactification of IIA on a
CY threefold \cite{DKV,KKV,KMV1,GMS,LMW} with a shrinking del Pezzo
surface in it. In this paper we will construct the precise map between
the spectra of the two different realizations of the theory.

We will first review both constructions. The states in the IIA theory
on the transverse space are obtained by wrapping D-branes on various
cycles in the del Pezzo and the description of the spectrum consists
of the classification of sheaves on the surface. On the Type IIB side
the states of the D3-brane world-volume theory are $(p,q)$ strings and
string junctions with support on 7-branes and the D3-brane. In the
next section we explain that both the image of the sheaves in the
K-theory group and the junctions form a lattice. In the third section
an isomorphism between these two lattices is established and in the
last section an application of this isomorphism is given which
identifies the Fourier-Mukai transform \cite{RP} of sheaves on ${\cal
B}_{9}$ with the $\sl2z$ symmetry of the dual brane configuration in
IIB.


\section{${\cal N}=2$ theories with $E_{N}$ global symmetry}

\subsection{Compactification of IIA}

We begin by reviewing how $d=4$, ${\cal N}=2$ field theories with
exceptional global symmetry can be be realized in IIA string theory.
Of the several ways involving the low energy limit of D-brane
configurations or compactifications we use the approach commonly
referred to as geometric engineering. Compactification of IIA string
theory on a non-compact CY threefold with a shrinking 4-cycle ${\cal
X}$ gives rise to a low energy theory with 8 supercharges on the
transverse space, whose spectrum is determined by the homology of that
4-cycle. When the shrinking manifold is a del Pezzo surface, the low
energy theory acquires exceptional global symmetry because the lattice
of 2-cycles contains the root lattice of an exceptional algebra and
thus admits the action of the corresponding Weyl group.

The states of the field theory are obtained from IIA D-branes wrapping
various submanifolds of ${\cal X}$: D0's on points, D2's wrapping
2-cycles and D4's on ${\cal X}$ itself. A configuration of $Q$
D4-branes is a $U(Q)$-bundle on ${\cal X}$ with instanton number given
by the number of D0's and first Chern class being the Poincare dual to
the homology class wrapped by D2-branes \cite{BSV,GD}. Proper description of these
D-brane configurations requires the more sophisticated notion of
sheaves on ${\cal X}$ \cite{MH,VW}. This becomes essential if
configurations with and without D4-branes are to be dealt with on
equal footing. 
BPS states arise from supersymmetric configurations of D-branes.
Wrapped D4-branes with immersed lower dimensional branes correspond to
semi-stable torsion-free sheaves on ${\cal X}$ \cite{MH}\footnote{In
  this case the sheaf is constructed from the sections of the
  holomorphic bundle.} while a generic configuration is described by a
coherent sheaf which is reducible to torsion and semi-stable
torsion-free components corresponding to non-immersed branes.

A del Pezzo surface ${\cal X}$ is a two complex dimensional manifold
constructed by blowing up $N$ points on $\PP^{2}$ or $N-1$ points on
$\PP^{1}\times \PP^{1}$, where $N \leq 8$. We denote these two
families as $\widetilde{\cal B}_{N}$ and ${\cal B}_{N}$, respectively;
moreover $\widetilde{\cal B}_{N}=\Bn$ for $N > 1$, and thus it is
sufficient to consider ${\cal B}_{1}=\PP^{1}\times \PP^{1}$ and
$\widetilde{\cal B}_{N}$ for $N=0\ldots 8$.  The almost del Pezzo
$\widetilde{\cal B}_{9}$ is also known as $\frac{1}{2}$K3 \cite{MNVW}
since it is an elliptically fibered manifold with a base $B\cong
\PP^{1}$ and twelve degenerate elliptic fibers.


\subsubsection{The homology of $\widetilde{\cal B}_{N}$ and the $E_N$ root lattice}

{\bf $\widetilde{\cal B}_{N\leq 8}$:} The 2nd homology group
$H_{2}(\widetilde{\cal B}_{N})$ is $N+1$ dimensional and is generated
by the elements $\{l,e_{1}, \ldots, e_{N}\}$, where $l$ is the
generator of $H_{2}(\PP^2)$ and $e_i\;(i=1 \ldots N)$ are the
exceptional curves. The intersection numbers are
\begin{eqnarray} 
\inter{l}{l}=1,\hspace{.3in} \inter{e_{i}}{e_{j}}=-\delta_{ij}, 
\hspace{.3in} \inter{l}{e_{i}}=0.
\end{eqnarray} 
The canonical class is $K_{\widetilde{\cal B}_{N}} = -3l+\sum_{i=1}^{N}e_{i}$, it is used
to define the degree of a 2-cycle $\Sigma$ as
\begin{eqnarray}
\mbox{d}_{\Sigma} \equiv -\inter{K_{\widetilde{\cal B}_{N}}}{\Sigma}, 
\;\;\;\;\;\Sigma\in H_{2}(\widetilde{\cal B}_{N}).
\end{eqnarray} 
The homology lattice $H_{2}(\widetilde{\cal B}_{N})$ contains the root
lattice $\Gamma_N$ of the $E_{N}$ algebra. To see this, identify the
set of roots as $\Delta_N = \{C\in H_{2}(\widetilde{\cal B}_{N},\bbbz) |
\inter{C}{C}=-2,
\mbox{d}_{C}=0\}$, and choose the simple roots of $E_N, \{3\leq N\leq
9\}$ as
\begin{eqnarray} 
C_{i}\equiv e_{i}-e_{i+1},~~i=1 \ldots N-1 \mbox{~~and~~} C_{N}
\equiv l- e_{1}-e_{2}-e_{3},
\label{roots}
\end{eqnarray} 
their intersection numbers yield the $E_N$ Cartan matrix.  Using $C_i$
as basis elements, we can define the weight vector
$\{\omega^{i}|i=1\ldots N\}$ and associate Dynkin labels $\lambda_{i}$ with each
curve:
\begin{eqnarray} 
\inter{\omega^{i}}{C_{j}}=-\delta^{i}_{j},~~
\lambda_{i} \equiv -\inter{C}{C_{i}},~~i=1\ldots N,
\end{eqnarray} 
such that the curve and the self-intersection is given in terms of the
corresponding weight vector:
\begin{eqnarray} 
\begin{array}{lllll}
\Sigma&=&\sum_{i=1}^{N}\lambda_{i}\omega^{i}-
\frac{\mbox{d}_{\Sigma}}{9-N}K_{\widetilde{\cal B}_{N}}\\  
\inter{\Sigma}{\Sigma}&=&
-\vec{\lambda}\cdot\vec{\lambda}+
\frac{\mbox{d}_{\Sigma}^{2}}{9-N},
\hspace{.8in}\Sigma\in H_2(\widetilde{\cal B}_{N\leq
8},\bbbz) . 
\end{array}
\end{eqnarray} 

{\bf ${\widetilde{\cal B}_{9}}\;$:} Blowing up one more point we
arrive at the final case we consider:$\widetilde{\cal B}_{9} =
\frac{1}{2}K3$ which is not strictly speaking del Pezzo. It is however
often referred to as almost del Pezzo having a nef (but not ample)
anticanonical class.
Since $\widetilde{\cal B}_{9}$ is elliptically fibered we will use a
different basis for $H_{2}(\widetilde{\cal B}_{9})$. We denote the
homology class of the base and the fiber by $B$ and $F$ respectively.
With the choice of basis $\{C_{1}\cdots C_{8}, B+F,B\}$,
$H_{2}(\widetilde{\cal B}_{9})=\Gamma_{E_{8}}\oplus\left({1\;\;
    \;\;0\atop \;0 \;\;-1}\right)$. In terms of the degree
$\mbox{d}_{\Sigma}$ and $\mbox{c}=\inter{B}{\Sigma}$ the curve
$\Sigma$ and its self-intersection number is given by
\begin{eqnarray} 
\begin{array}{lllll}
\Sigma&=&\sum_{i=1}^{8}\lambda_{i}\omega^{i}+\mbox{d}_{\Sigma}(B+F)+
\mbox{c}F \\
\inter{\Sigma}{\Sigma}&=&
-\vec{\lambda}\cdot\vec{\lambda}+\mbox{d}^{2}_{\Sigma}+
2\mbox{c}\mbox{d}_{\Sigma}, 
\hspace{.8in}\Sigma\in H_{2}(\widetilde{\cal B}_{9},\bbbz).
\end{array}
\end{eqnarray} 
Since $H_{2}(\widetilde{\cal B}_{9})$ contains the affine $E_{8}$ root
lattice, it admits the action of the affine $E_{8}$ Weyl group and all
curves fall into representations of affine $E_{8}$ such that
$\mbox{d}_{\Sigma}$ and $\mbox{c}$ are the level and the grade of the
representation respectively.

\subsubsection{The K-theory group of del Pezzo surfaces}

The K-theory group of $\widetilde{\cal B}_{N}$ and ${\cal B}_{1}$ are
given as \cite{KN}:
\begin{eqnarray} 
 \mbox{K}(\widetilde{\cal B}_N)=
\underbrace{\bbbz\oplus \cdots \oplus \bbbz}_{N+3},
\hspace{0.5in}
\mbox{K}({\cal B}_{1})=\bbbz\oplus \bbbz\oplus \bbbz \oplus \bbbz \,.
\end{eqnarray} 
The K-theory group $K^0$ of $\widetilde{\cal B}_{N} ~({\cal B}_{1})$
is generated by $N+3~ (4)$ elements and being torsion free it is
isomorphic to the even-degree integral cohomology group 
$H^{*}(\widetilde{\cal B}_{N}) ~(H^{*}({\cal B}_{1}))$.
The map is given by
\begin{eqnarray} 
\mbox{K}({\cal X})\ni{\cal E}\longrightarrow \mbox{ch}({\cal E})\in
H^{*}({\cal X}).
\end{eqnarray} 
There is a natural bilinear form on $\mbox{K}({\cal X})$ \cite{KN}: let ${\cal
E}_{1,2}\in \mbox{K}({\cal X})$ then
\begin{eqnarray} 
\langle{\cal E}_{1},{\cal E}_{2}\rangle &\equiv&
\int_{X}\mbox{ch(${\cal E}_{1} \otimes {\cal E}^{*}_{2}$)}\wedge\mbox{Td}({\cal X})
\nn\\ &=&\int_{X}\mbox{ch(${\cal E}_{1})$}\wedge\mbox{ch(${\cal
E}^{*}_{2}$)}\wedge\mbox{Td}({\cal X}) \equiv\int_{X}\mbox{ch(${\cal
E}_{1})$}\wedge\mbox{ch(${\cal E}_{2}$)}^{\vee}\wedge \mbox{Td}({\cal X}),
\end{eqnarray} 
where ${\cal E}^{*}$ is the dual bundle, $\mbox{Td}({\cal X})= 1+\frac{1}{2}c_{1}({\cal
X})+\frac{1}{12}(c_{1}({\cal X})^{2}+c_{2}({\cal X}))$ and if
$v=\sum_{i=0}^{2}v_{i},\, v_{i}\in H^{2i}({\cal X})$ then $v^{\vee}
\equiv \sum_{i=0}^{2}(-1)^{i}v_{i}$.  Let us write\footnote{We will
use the same symbol for the 2-form and its dual 2-cycle. Thus
$\inter{\Sigma_{a}}{\Sigma_{b}}\equiv \int_{{\cal X}} \Sigma_{a}\wedge
\Sigma_{b}.$} the Chern classes of ${\cal E}_{1,2}\in \mbox{K}({\cal
X})$ as ch$({\cal E}_{a})=(\mbox{r}_{a}, \Sigma_{a},
\mbox{ch}_{2}({\cal E}_{a}))$; then we obtain\footnote{$\int_{{\cal
X}}c_{1}({\cal X})\wedge c_{1}({\cal X})=9-N$ and $\int_{{\cal
X}}c_{2}({\cal X})=N+3$ for ${\cal X}=\tilde{\cal B}_{N}$\,.}
\begin{eqnarray} 
\langle{\cal E}_{1},{\cal E}_{2}\rangle = \mbox{r}_{1}\mbox{r}_{2}-
\inter{\Sigma_{1}}{\Sigma_{2}}+\mbox{r}_{1}\mbox{ch}_{2}({\cal E}_{2})
+\mbox{r}_{2}\mbox{ch}_{2}({\cal E}_{1})+
\frac{1}{2}(\mbox{r}_{2}\mbox{d}_{\Sigma_{1}}-
\mbox{r}_{1}\mbox{d}_{\Sigma_{2}}).
\end{eqnarray} 
An element $(\mbox{r},\,\Sigma,\,\mbox{ch}_2)\in H^{*}({\cal X})$
represents an equivalence class ${\cal E}$ of sheaves on ${\cal X}$
such that ch(${\cal E})$=(r, $\Sigma$, ch$_{2}({\cal E})$) where
$\mbox{ch}_{2}({\cal E})=
\frac{1}{2}\inter{\Sigma}{\Sigma}-\int_{\cal X}c_{2}({\cal E})$.

Note that this bilinear form is not symmetric, the antisymmetric
part is given by the determinant
\begin{eqnarray} 
\langle{\cal E}_{1},{\cal E}_{2} \rangle-\langle{\cal E}_{2},{\cal
E}_{1}\rangle= 
-\left|\begin{array}{cc}
\mbox{r}_1          & \mbox{r}_2 \\
\mbox{d}_{\Sigma_1} & \mbox{d}_{\Sigma_2}
\end{array}\right|.
\end{eqnarray} 
With this bilinear form $K(\widetilde{\cal B}_{N})$ has signature
$(N+1,2)$. This scalar product was also considered in \cite{zaslow}
where it was shown that for certain manifolds the matrix of scalar
products of exceptional sheaves is the same as the soliton counting
matrix \cite{BCV} of the corresponding Landau-Ginzburg theory. We plan
to explore the relation between the exceptional sheaves on del Pezzos,
solitons of the corresponding ${\cal N}=2$ Landau-Ginzburg theories
and string junctions living on affine 7-brane backgrounds further in 
the future.


\subsection{Probe theory in IIB}

Four-dimensional ${\cal N}=2$ theories can be realized as the
world-volume theory of a D3-brane probe in the vicinity of some
7-branes of IIB string theory. The algebra of the 7-branes
appears as the global symmetry in the 4d field theory. The states
arise from strings and string junctions\cite{witten/aharony/schwarz}
stretched between the D3 and (some of) the 7-branes and are
characterized by the charges of each 7-brane.  Viewing the setup as an
F-theory compactification on K3, these states have the following
geometrical interpretation. The position of the D3-brane singles out a
fixed elliptic fiber $E_{*}$ and the strings/junctions with a
$({p\atop q})$ string segment ending on the D3-brane can be viewed as
curves in the K3 whose boundary wraps the $(p,q)$-cycle of $E_{*}$. In
the dual M-theory picture the selected elliptic fiber is wrapped by an
M5-brane and the states arise from M2 branes wrapping the curves in
K3 which end on the M5-brane \cite{MNS}.

The low energy field theory on the D3-brane in the vicinity of certain
configurations of $m$ 7-branes is insensitive to the remaining $24-m$
7-branes \footnote{It was shown in \cite{SZ} that the only 7-brane
backgrounds which allow such a decoupling are the ones with elliptic
or parabolic monodromy.}. For the sake of simplicity we might replace
the K3 in our arguments by an elliptically fibered {\em noncompact}
Ricci-flat manifold ${\cal M}_m$ containing these $m$ singular fibers
only.  Then the spectrum of the four dimensional probe theory is
characterized by the curves of (the elliptically fibered) ${\cal M}_m$
with boundary on the selected elliptic fiber $E_{*}$, or the relative
homology $H_2({\cal M}_m,E_{*})$.  The natural norm on this lattice is
given by the self-intersection of the homology elements. The BPS
states correspond to the holomorphic curves, which satisfy
\begin{eqnarray}
\inter{\C}{\C}=2g-2+b,
\end{eqnarray}
$b$ and $g$ being the number of boundary components on the $E_{*}$ and
the genus of the curve respectively.

Let us now specify the particular 7-brane backgrounds of our
interest. The well-known ${\bf E_{N}}$ Kodaira singularities ($N< 9$)
can be understood in terms of $N+2$ 7-branes, of which at
most $N$ may be mutually local. A string (junction) state can be
identified by specifying its charge with respect to each 7-brane
(linking number \cite{HW} or invariant charge \cite{DZ}) and thus is
an element of an $N+2$ dimensional lattice. To find the theories which
are dual to the ones presented in the previous section, we need a
lattice of one more dimension. As it was pointed out in \cite{DHIK},
the relevant 7-brane background is obtained by adding an extra 7-brane
to the ${\bf E_N}$ configuration so that it becomes ${\bf
\widehat{E}_N}$, which was studied in detail in \cite{dewolfe,part1,part2} and
is summarized in the following subsections.  The junction lattice of
${\bf \widehat{E}_{N}}$, ${\bf {\cal J}}^{2,N+1}$, is $N+3$
dimensional and is of signature $(2,N+1)$. 

\subsubsection{${\bf \widehat{E}_{N< 9}}$} 
We consider the type IIB background with $N+3$ non-local 7-branes of the 
configuration ${\bf \widehat{E}}_{N<9}$. 
The elliptic fibration of the corresponding F-theory manifold, $\Eman$ is
characterized by the monodromy around the singular fibers which is
encoded in the 7-brane charges \cite{GHZ}. Denote a $[p,q]$ 7-brane as
${\bf X_{[p,q]}}$, with corresponding inverse monodromy\footnote{
$K_{p,q}$ is the $\sl2z$ action felt by a string as it crosses the
branch cut of the 7-brane.}
$K_{p,q}=\left({
1+pq\;\; -p^2\atop \;\;\;q^2 \;\;\;\;1-pq}\right)$. Then 
$\Eman$ is defined in terms of the following 7-brane
configuration\footnote{This configuration is identical to ${\bf
\widehat{\tilde{E}}_N}$ of \cite{part2} up to an overall
transformation with $T^4\in\sl2z$.}: 
\begin{eqnarray}
\label{conf}
{\bf (X_{[1,0]})^{N}\;X_{[6,-1]} X_{[-3,1]}\;\;\;
X_{[0,1]}}\hspace{.8in}
K(\Eman)=\pmatrix{1 & 9\!-\!\!N \cr 0 & 1 },
\end{eqnarray}
$K(\Eman)$ being the overall monodromy.  If we remove the ${\bf
X_{[0,1]}}$-brane, the remaining \mbox{$N\!+\!2$} 7-branes can be
collapsed to the ${\bf E_N}$ Kodaira-singularity with overall
monodromy and associated binary quadratic form \cite{part1}:
\begin{eqnarray} 
K({\cal E}_{N})=\pmatrix{1 &9-N \cr -1 & N-8 } \hspace{.5in}
f_{E_{N}}(p,q)=\frac{p^{2}}{9-N}+pq+q^2.
\end{eqnarray} 
Let us consider ${\bf E_N}$ first. In a suitable basis
\cite{DZ} a junction can be written as
$\mJ=\sum_{i=1}^{N}\lambda_{i}\momega^{i} + p\momega^{p} +
q\momega^{q}$, and the self-intersection form on the homology lattice
$H_2({\cal E}_N,E_{*})$ factorizes such that the curve $\mJ$ with
boundary wrapping the $(p,q)$-cycle of the $E_{*}$ has norm
\begin{eqnarray}
(\mJ,\mJ)=-\lambda^{2}_{E_N}+f_{E_{N}}(p,q),
\label{ennorm}
\end{eqnarray}
where $\lambda$ is a vector of the $E_N$ weight lattice. Back to
$\Eman$, the addition of the ``affining'' ${\bf X_{[0,1]}}$ 7-brane
gives rise to one more basis element, which we call ${\bf \delta}^{(-1,0)}$ and
choose to be a closed $({-1\atop 0})$ string (oriented
counterclockwise) encircling all of the 7-branes:
\begin{eqnarray}
\mJ&=&\sum_{i=1}^{N}\lambda_{i}\momega^{i}+p\momega^{p}+
                    q\momega^{q}+n{\bf \delta}^{(-1,0)}, \\
(\mJ,\mJ)&=&
-\lambda^{2}_{E_{N}}+2nq+f_{E_{N}}(p,q).
\label{selfint}
\end{eqnarray}
In this case the norm does not factorize as in \myref{ennorm}; the
first two terms, however, can be regarded as the norm of a weight
vector in the affine lattice $\widehat{E}_N$. 

Eqns. \myref{ennorm} \myref{selfint} determine the self-intersection
number of a curve in $\Eman$ with boundary on the selected
fiber. There is some ambiguity in extending this formula to mutual
intersection between curves whose boundary wraps different cycles of
the torus, which is resolved by specifying the contribution of the
intersecting boundaries of two curves depicted in \figref{mutual}.  
\onefigure{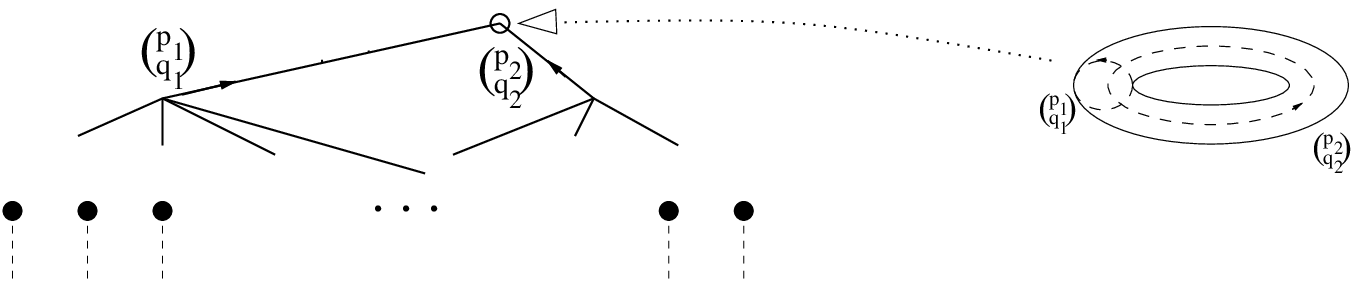}{ Two curves (junctions) whose boundary wraps
intersecting cycles of the selected elliptic fiber.}
Linearity of the intersection number requires that if $\mJ_1$ and
$\mJ_2$ wrap the $(p_1,q_1)$ and $(p_2,q_2)$ cycle of the $T^2$
respectively, then the contribution from the boundary is
\begin{eqnarray}\mbox{$
(\mJ_1,\mJ_2)_{boundary} = \alpha \left|{p_1\;\;p_2\atop
q_1\;\;q_2}\right|
\hspace{.5in} 
(\mJ_2,\mJ_1)_{boundary} = (1-\alpha) \left|{p_1\;\;p_2\atop
q_1\;\;q_2}\right|.
$}
\label{boundterm}
\end{eqnarray}
In \cite{DZ} the contribution was postulated to be
symmetric ($\alpha=\half$), which led to fractional intersection
numbers in general. This is not suitable for us because the metric on
the lattice of the dual theory is manifestly integral. We shall
utilize the simplest integral choice: $\alpha=1$, so that
$(\mJ_1,\mJ_2)_{boundary} = p_1q_2-q_1p_2$ and
$(\mJ_2,\mJ_1)_{boundary} = 0$. Together with \myref{selfint} this
uniquely fixes the intersection matrix on the junction lattice which
is straightforward to determine. 

{\bf Summary}: A 3-brane parallel to the singular fibers of an
F-theory compactification on a $\Eman$ manifold realizes a $d=4$,
${\cal N}=2$ theory whose states are characterized by an $N+3$
dimensional charge lattice equipped with following intersection bilinear
form in the basis $\{\alpha_{i}|i=1,...,N; \omega^{p}\,,\, \omega^{q}\,,\,\delta^{(-1,0)}\}$:
\begin{eqnarray}
\left(\begin{array}{cccc}
-E_N&&& \\
&\frac{1}{9-N}&1&0 \\
&0&1&1 \\
&0&1&0
\end{array}\right),
\end{eqnarray}
where $E_N$ is the Cartan matrix of the corresponding Lie algebra.

\subsubsection{${\bf \widehat{E}_{9}}$}
Going beyond $N=8$ in the series of the 7-brane configurations of
\myref{conf} we encounter $\widehat{\cal E}_9$ which is special from
numerous aspects. The overall monodromy is trivial and as a
consequence, it admits two linearly independent closed strings
encircling the 7-branes. It is useful to visualize the configuration
in the following way:
\begin{eqnarray}
{\bf X_{[1,0]}\;\;\;
\left((X_{[1,0]})^8\;X_{[6,-1]} X_{[-3,1]}\right)
\;\;\;X_{[0,1]}}
\;\;\;\; = \;\;\;\;
{\bf X_{[1,0]}\;\;\;
\left(E_8\right)
\;\;\;X_{[0,1]}},
\end{eqnarray}
and think about it as being ``doubly affined'' \cite{part2}. One
possible basis for the lattice of junctions being supported on this
configuration is 
\begin{eqnarray}
\mJ&=&\sum_{i=1}^8\lambda_{i}\momega^{i}+p\momega^{p}+q\momega^{q}+
n_1\,{\bf \delta}^{(-1,0)}+ n_2\,{\bf \delta}^{(0,1)} \\
(\mJ,\mJ)&=& -\lambda^{2}_{E_8}+2n_1q+2n_2p+f_{E_8}(p,q) = \nn\\
&=&-\lambda^{2}_{E_8}+2n_1q+2n_2p+p^2+pq+q^2,
\end{eqnarray}
with intersection matrix in the basis
$\{\alpha_{i}|i=1,..,8\,;\,\omega^{p}\,,\,\omega^{q}\,,\,\delta^{(-1,0)}\,,\,\delta^{(0,1)}\}$ is
\begin{eqnarray}
\left(\begin{array}{ccccc}
E_8&&&& \\
&1&1&0&1 \\
&0&1&1&0 \\
&0&1&0&0 \\
&1&0&0&0
\end{array}\right).
\end{eqnarray}

\subsection{3-cycles and string junctions}

Mirror symmetry relates type IIA string theory compactified on a
Calabi-Yau threefold $M$ to type IIB on the mirror manifold $W$
\cite{mirror}. It follows from the interpretation
of mirror symmetry as T-duality \cite{SYZ,LV} that the even cohomology
classes of $M$ are mapped to the odd cohomology classes of $W$ and
therefore the complexified K\"ahler structure parameters of $M$ are
exchanged with the complex structure parameters of $W$ \cite{morrison}.
Vector bundles with characteristic classes represented by 
the even cohomology classes map to 3-cycles dual to the odd cohomology 
classes \cite{vafa-bundle}.  
\subsubsection{The mirror of a Calabi-Yau threefold containing 
${\cal B}_{9}$}

A Calabi-Yau threefold, $M$, containing ${\cal B}_{9}$ can be
described as a double elliptic fibration over $\PP^{1}$
\cite{MNVW}, 
\begin{eqnarray} 
y_i^2 = x_i^3 + f_{4,i}(z)\,x_i+g_{6,i}(z), \hspace{.4in} i=1,2,
\end{eqnarray} 
where $z$ is the coordinate on the $\PP^{1}$. The total space of each
fibration over the sphere defines a ${\cal B}_{9}$ surface. As shown
in \cite{MNVW}, this CY threefold can also be obtained by resolving
the singularities of a $\bbbz_2\times \bbbz_2$ orbifold of $T^2 \times
T^2\times T^{2}$. If $\xi_{1,2,3}$ are the complex coordinates on the
three tori then the orbifold action is given by
\cite{MNVW}
\begin{eqnarray} 
(\xi_{1},\xi_{2},\xi_{3})\rightarrow 
(\xi_{1}+\shalf,\xi_{2},-\xi_{3}) 
\hspace{.3in}\mbox{and}\hspace{.3in}
(\xi_{1},\xi_{2},\xi_{3})\rightarrow 
(\xi_{1},\xi_{2}+\shalf,-\xi_{3}),
\end{eqnarray} 
and the third torus becomes the $\PP^1$ after the identification. The
holomorphic 3-form is \mbox{$\Omega^{(3)} = dz\frac{dx_1}{y_1}
\frac{dx_2}{y_2}$} with $z$ being the coordinate on the $\PP^1$. To
obtain the non-compact CY-threefold containing ${\cal B}_{9}$, we
decompactify the fiber of one of the elliptic fibrations by taking its
area to infinity i.e. we consider the limit of the complexified
K\"ahler parameter \mbox{$B+iA \equiv \rho\rightarrow i\infty$}.

The mirror threefold $W$ is obtained by performing T-duality on one of
the cycles of each torus. The decompactification limit of $M$ then maps
to the limit of $W$ when the complex structure parameter $\tau$ of the
elliptic fibration goes to $i\infty$ which in effect decompactifies
one of the cycles of the torus. The local model for this is given by
\begin{eqnarray} 
y_{2}^{2}-x_{2}^{2}=(z-z_{*}).
\end{eqnarray}  
Since we want the degenerate complex structure limit for the entire
elliptic fibration, we adopt the above model globally over
$\PP^{1}$. The structure of $W$ then is that of a $T^2\times S_{c}^1\times
I\!\!R$ fibration over a $\PP^{1}$ and the holomorphic 3-form becomes
\begin{eqnarray} 
\Omega^{(3)} = dz\frac{dx_1}{y_1} \frac{dx_2}{x_2} =
\frac{dx_2}{x_2}\Omega^{(2)},
\end{eqnarray} 
where $\Omega^{(2)}$ is the holomorphic 2-form on ${\cal
B}_{9}$. Since the canonical bundle of ${\cal B}_{9}$ is non-trivial
this holomorphic two form has zeros or poles at the 2-cycle dual to
the first Chern class of the canonical bundle. The total space can be
visualized as a double fibration over $\PP^1$: the $T^2$ fibration
constitutes a ${\cal B}_9$ surface and a cylinder is also fibered over
its base. At one point $z_{*}\in \PP^{1}$ the nontrivial cycle of the
$S_{c}^{1}\times I\!\!R$ shrinks to zero size. We denote the elliptic
fiber of ${\cal B}_9$ above this point by $E_{*}$. 
\onefigure{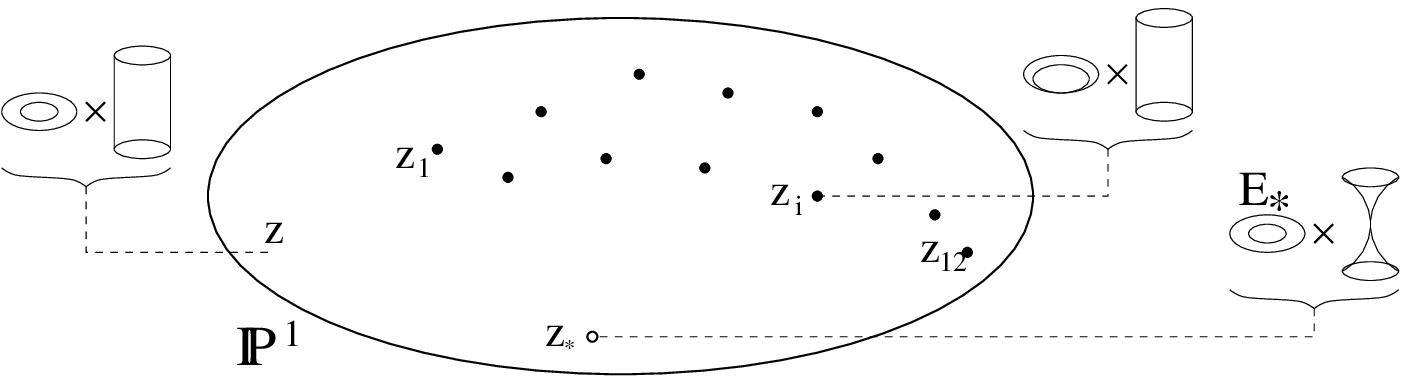}{The non-compact mirror Calabi-Yau $W$, as an $S_{c}^1\times I\!\!R$
fibration over the base of the elliptically fibered ${\cal B}_9$. At
the points $z_i$ different cycles of the elliptic fiber of ${\cal
B}_9$ are shrinking while at $z_*$ the $S_{c}^1$ shrinks.}

\subsubsection{3-Cycles in $W$} 

Mirror symmetry maps the even homology of $M$ to the odd homology of
$W$. Since $W$ is simply connected the only odd homology elements of
$W$ are the 3-cycles of the following type \cite{MNVW}:
\begin{itemize}
\item $S_{c}^1\times S^2$: Here $S^2=C$ is a curve in ${\cal B}_9$ and
$S_{c}^1$ is the non-trivial cycle of the $S^1\times I\!\!R$ 
fibration. There are eight such cycles on which the holomorphic 3-form
$\Omega^{(3)}$ is non-zero: \mbox{$S_{c}^{1}\times C_{i}$},
\mbox{$i=1\ldots 8$} with $C_i$ given in \myref{roots} 
corresponding to the roots of the $E_8$ root lattice embedded in
$H_2({\cal B}_9)$. 
\item $S^3$: There are two such (linearly independent) cycles. They
are formed by $S_{c}^1\times S^1$ fibered over intervals ${\cal I}_{1,2}$
such that $S^1$ shrinks on one end of the interval and
$S_{c}^{1}$ shrinks at the other side. 
\item $T^{3}$: There are again two (linearly independent) cycles of 
this type. These are formed by $S_{c}^{1}$ in the $S_{c}^{1}\times
I\!\!R$ fibration and a torus in the ${\cal B}_{9}$. The torus in the
${\cal B}_{9}$ is formed from the circle surrounding the position of
the degenerate fibers on the base and a 1-cycle of the elliptic fiber. 
\end{itemize}

\subsubsection{String junctions from 3-cycles} 
Let $\pi:{\cal B}_{9}={\cal X}\rightarrow \PP^{1}$ be an elliptic
fibration.  We denote by $E_{*}=\pi^{-1}(z_{*})$, as discussed before,
a fixed non-degenerate elliptic fiber.  The section
$e:\PP^{1}\rightarrow {\cal X}$ is such that $e(\PP^{1})=B$ with
$B^{2}=-1$. The three-cycles in the non-compact Calabi-Yau threefold
$W$ are of the form ${\cal C}\times S^{1}_{c}$, where ${\cal C}$ is a
curve in ${\cal B}_{9}$ with a boundary such that ${\cal C}\cap
E_{*}=\partial {\cal C} \in H_{1}(E_{*},\bbbz)$.  With any such curve
${\cal C}\in {\cal X}$ we can associate a junction $\mJ_{\cal C}$
living on the base $B$,
\be
\mJ_{\cal C}=e(\pi({\cal C})).
\ee 
If $\partial {\cal C}=0$, the corresponding junction has no asymptotic
charge and has support only on the 7-branes, the positions of the 
degenerate fibers.  When $\partial {\cal C}=(p,q)\in H_{1}(E_{*},\bbbz)$ 
the corresponding junction has asymptotic charge
$(p,q)$ on the D3-brane, the position of $E_{*}$ on the base. 

We summarize the mirror symmetry map between the homology of ${\cal
B}_{9}$ and the 3-cycles of the mirror Calabi-Yau $W$ in the following
table. ${\cal I}_{1}$ and ${\cal I}_{2}$ represent curves on the
base from $z_{1}$ and $z_{12}$ to $z_{*}$ respectively.  $\delta$ is
the path encircling the 7-branes.  $S^{1}$ and $S^{1}_{D}$ are the two
basis 1-cycles of the elliptic fiber.
\begin{eqnarray}
\begin{array}{|c|rcl|c|rcl|}
\hline 
\rule{0mm}{6mm}{\bf H_{*}({\cal B}_{9})}  & \multicolumn{3}{c|}{\bf H_{3}(W)} 
&{\cal J}^{2,10} & A(\Sigma)&=&\int_{\Sigma}\Omega^{(2)}
\\ \hline \hline
\rule{0mm}{6mm}C_{i},\;\mbox{\footnotesize $i=1\ldots8$} & S^2\times S_{c}^1&=&C_{i}\times S_{c}^{1} & 
C_{i} & m_i&=&\int_{C_{i}}dz~\frac{dx}{y}
\\ \hline
\rule{0mm}{6mm}B               & S^3&=&{\cal I}_{1}\times (S^1\times S_{c}^1) & 
{\bf a}_{1} ={\bf x}_{[1,0]}   & \phi&=&\int_{z_1}^{z_*}dz\oint_{[1,0]}\frac{dx}{y}
\\ \hline
\rule{0mm}{6mm}F               & T^3&=&\delta \times S^1\times S_{c}^1          &  
{\bf \delta}^{(0,1)} & \tau&=&\oint_{\delta}dz\oint_{[0,1]}\frac{dx}{y}
\\ \hline
\rule{0mm}{6mm}{\cal B}_{9}    & S^3&=&{\cal I}_{2}\times(S^1_D\times S_{c}^1) & 
{\bf x}_{[0,1]} &
\phi_{D}&=&\int_{z_{12}}^{z_*}dz\oint_{[0,1]}\frac{dx}{y}
\\ \hline
\rule{0mm}{6mm}\mbox{0--cycle} & T^3&=&\delta \times S_{D}^1\times S_{c}^1  &
{\bf \delta}^{(-1,0)}  & 
\tau_{D}&=&\oint_{\delta}dz\oint_{[-1,0]}\frac{dx}{y}\\ 
\hline
\end{array}
\nonumber
\end{eqnarray}

The following table shows the del Pezzo surfaces and the corresponding
dual 7-brane configurations:
\begin{eqnarray} 
\begin{array}{|rcl|l|c|} 
\hline
\mbox{Complex}&&\mbox{surface} & \mbox{Brane Configuration} &  \mbox{Algebra} 
\\ \hline \hline
\rule{0mm}{6mm} \tilde{\cal B}_{0}&=&\PP^{2}& 
{\bf X_{[6,-1]}X_{[-3,1]}X_{[0,1]}} &  \widehat{E}_{0} 
\\ \hline
\rule{0mm}{6mm} \widetilde{\cal B}_{1}&=& \PP^{2}\#\bar{\PP}^2 & 
{\bf X_{[6,-1]}X_{[-3,1]}X_{[0,1]}} & \widehat{\tilde{E}}_1=\widehat{u(1)} 
\\ \hline
\rule{0mm}{6mm} {\cal B}_{1}&=&\PP^{1}\times \PP^{1} & 
{\bf X_{[1,-1]}X_{[1,1]}X_{[1,-1]}X_{[1,-1]}} &
\widehat{E}_1=\widehat{su(2)} 
\\ \hline
\rule{0mm}{12mm}\widetilde{\cal B}_{N>1}&=&\left\{\begin{array}{ll}
\PP^{2}\#\underbrace{\bar{\PP}^2\#\ldots
\#\bar{\PP}^{2}}_{N}\\ 
(\PP^{1}\times\PP^{1})\#
\underbrace{\bar{\PP}^2\#\ldots
\#\bar{\PP}^{2}}_{N-1}\end{array}\right.  & 
\left\{\begin{array}{ll}{\bf A}^{N}{\bf X_{[6,-1]}X_{[-3,1]}X_{[0,1]}}\\  ~\\ 
{\bf A}^{N-1}({\bf X_{[1,-1]}X_{[1,1]})^{2}}\end{array}\right.
 &
\widehat{E}_{N} 
\\ \hline
\end{array}\nn
\end{eqnarray} 

\section{String junctions and Vector bundles}

\subsection {${\cal B}_9$ and ${\bf \widehat{E}_9}$}

We map the equivalence classes of sheaves on ${\cal B}_{9}$ to string
junctions with support on the $\widehat{\bf E}_9$ 7-brane
configuration.  Recall that in \cite{DHIK} the map between curves
and junctions of zero $q$ charge living on ${\bf E_{9}}$ 7-brane
configuration was given. According to that map a curve
\be
\Sigma= \sum_{i=1}^{8}\lambda_{i}\omega^{i}+
\mbox{d}_{\Sigma}(B+F)+\mbox{c} F \in H_{2}({\cal B}_{9},\bbbz),
\ee
corresponds to a family of junctions
\be
\mJ_{\Sigma}(m)\equiv\sum_{i=1}^{8}\lambda_i\momega^{i}+
\mbox{d}_{\Sigma}\momega^p+
\mbox{c}\delta^{(0,1)}+m\delta^{(-1,0)}\,,\hspace{.4in}m\in\bbbz\,.
\ee
Different values of $m$ correspond to different bundles on $\Sigma$.
We denote by ${\cal O}_{\Sigma}(m)$ a (torsion sheaf whose restriction
to its support is a) bundle on $Sigma$ with $\mbox{ch}({\cal
  O}_{\Sigma}(m)) = (0\,,\,\Sigma\,, \,m+\frac{1}{2}
\mbox{d}_{\Sigma})$. 
A D2-brane wrapped on $B$ maps to a D4-brane wrapped on ${\cal B}_{9}$
after T-duality, which we will interpret as an $\sl2z$ transformation
by $S$.  Therefore we require that the map between the junctions and
bundles should satisfy the following conditions:
\begin{itemize}
\item The rank of the bundle ${\cal F}$ is the $q$-charge of the 
corresponding junction.
\item The degree $\mbox{d}_{\Sigma}$ of the first Chern class
$\Sigma$ of the bundle, is the $p$-charge of the junction.
\item $\langle {\cal F}, {\cal F}\rangle =- \inter{\mJ_{\cal
F}}{\mJ_{\cal F}}$.
\end{itemize}
By equating the two scalar products and requiring that they be equal
for all $(r, \mbox{d}_{\Sigma})=(q, p)$ we get a unique map between the bundle data and the junction data.
It follows that a bundle ${\cal F}$ with
\be
\mbox{ch}({\cal F})=(\mbox{r}\,,\Sigma\,, \mbox{k})\,,\hspace{.5in}
\Sigma=\sum_{i=1}^{8}\lambda_{i}\omega^{i}+
\mbox{d}_{\Sigma}(B+F)+\mbox{c} F,  
\ee
corresponds to the junction
\begin{eqnarray}
\begin{array}{llllcl}
\mJ_{\cal F}=\sum_{i=1}^{8}\lambda_{i}\momega^i+
\mbox{d}_{\Sigma}\momega^p+c {\bf \delta}^{(0,1)}
+\mbox{r}\momega^q-(\mbox{r}+\mbox{k}+
\frac{1}{2}\mbox{d}_{\Sigma}){\bf \delta}^{(-1,0)}.
\label{map}
\end{array}
\end{eqnarray}

\subsection{$\Bn$ and ${\bf \widehat{E}_{N}}$} 

The map between $K(\Bn)$ and ${\bf \widehat{E}_{N}}$ for $N<9$ follows
from the above map.  Blowing down an exceptional curve corresponds to
removing a ${\bf X_{[1,0]}}$ 7-brane of the ${\bf \widehat{E}_9}$
7-brane configuration.  By this process we not only decouple the string
junction with support on that brane but we also lose the
$\delta^{(0,1)}$ string junction. It then follows from \myref{map}
that a class ${\cal F}\in K(\Bn)$ such that
\be
\mbox{ch}({\cal F})=(\mbox{r}\,,\Sigma\,,\mbox{k})\,,\,\,\,\
\Sigma=\sum_{i=1}^{N}\lambda_{i}\omega^{i}-\frac{\mbox{d}_{\Sigma}}{9-N}K_{\Bn}\in
H_{2}(\Bn,\bbbz),
\ee
corresponds to  
\be
\mJ_{\cal F}=\sum_{i=1}^{N}\lambda_{i}\omega^{i}+\mbox{d}_{\Sigma}\omega^{p}
+\mbox{r}\omega^{q}-\{\mbox{r}+\mbox{k}+\frac{1}{2}\mbox{d}_{\Sigma}\}\delta^{(-1,0)}\,.
\ee 
where $\mbox{r} +\mbox{k}+\frac{1}{2}\mbox{d}_{\Sigma}\equiv \chi({\cal F}) $ 
is the Euler-Poincare characteristic
of ${\cal F}$.  $K(\Bn)$ is an abelian group, the inverse of ${\cal
F}$ is $-{\cal F}$ and the associated junction is $-\mJ_{\cal F}$.  With this
identification we get
\be
\langle {\cal F}_{1}, {\cal F}_{2} \rangle =- \inter{\mJ_{{\cal
F}_{1}}}{\mJ_{{\cal F}_{2}}}. 
\ee

\subsection{Genus of the junction and dimension of the moduli space} 

Let ${\cal F}$ be a stable holomorphic vector bundle on ${\cal B}_{9}$
with $\mbox{ch}({\cal F})=(\mbox{r},\Sigma,\mbox{ k})$ such that
gcd$(\mbox{r},\mbox{d}_{\Sigma})=1$.  We denote the corresponding
special Lagrangian 3-cycle in the mirror Calabi-Yau and the BPS
junction by $C_{\cal F}$ and $\mJ_{\cal F}$ respectively.  The
dimension of the moduli space ${\cal M}(\mbox{r},\Sigma,\mbox{k})$ of
the vector bundle is $-\langle {\cal F}, {\cal F} \rangle+1$ while the
moduli space is empty if $\langle {\cal F}, {\cal F} \rangle > 1$
\cite{HL,bridgeland}.  From the correspondence with junctions
we see that \cite{ADE-paper}
\begin{eqnarray}
\begin{array}{llllcl}
dim~{\cal M}(\mbox{r},\Sigma,\mbox{k})&=&
\inter{\mJ_{\cal F}}{\mJ_{\cal F}}+1\\
&=&2g-2+\mbox{gcd}(\mbox{r},\mbox{d}_{\Sigma})+1=2g,
\end{array}
\end{eqnarray}
where $g$ is the genus of the curve associated with the junction.
This is in agreement with Vafa's conjecture \cite{vafa-bundle} that
the map between the 3-cycles and the vector bundles should be such
that
\be
H^{i}(\mbox{End} ~{\cal F})=H^{i}(C_{\cal F},W).
\ee
It follows from the above identification that
\begin{eqnarray}
\begin{array}{llllcl}
\langle {\cal F}, {\cal F}\rangle &\equiv&\sum_{i=0}^{2}(-1)^{i}dim~ 
\mbox{Ext}^{i}({\cal F},{\cal F})
\equiv \sum_{i=0}^{2}(-1)^{i}dim ~H^{i}(\mbox{End}~ {\cal F})\\
&=& \sum_{i=0}^{2}(-1)^{i}dim ~H^{i}(C_{\cal F},W)= 
\sum_{i=0}^{2}(-1)^{i}b^{i}\\
&=&\chi(C_{\cal F})+1
\end{array}
\end{eqnarray}
and therefore 
\begin{eqnarray}
\begin{array}{llllcl}
dim~{\cal M}(\mbox{r},\Sigma,\mbox{k})&=&-\langle {\cal F},{\cal F}\rangle+1\\
&=&-\chi(C_{\cal F})=2g.
\end{array}
\end{eqnarray}
We summarize the results of this section in the following
table. ${\cal O}_{X}$ is the structure sheaf of the manifold $X$ and
is the trivial rank one bundle corresponding to a D4-brane wrapped on
$X$. ${\cal O}_{\Sigma}(m)$ and ${\cal O}_{x}$ are the torsion sheaf
and the skyscraper sheaf respectively.
\begin{eqnarray} 
\begin{array}{|c|c|c|c|}
\hline
\mbox{D-branes}&{\cal F} &  \mbox{ch}({\cal F})& \mbox{string junction} 
\\  \hline
\mbox{D4-brane}& \rule{0mm}{5mm}{\cal O}_{\widetilde{\cal B}_{N}} 
 & (1,0,0) &\momega^{q}-\delta^{(-1,0)} \\ \hline 
\mbox{D2-brane}+m\mbox{D0-branes}&\rule{0mm}{6mm}{\cal
O}_{\Sigma}(-m), 
m\in\bbbz_{\geq 0} &(0,\Sigma,\frac{1}{2}\mbox{d}_{\Sigma}-m)
 &\mJ_{\Sigma}(m-\mbox{d}_{\Sigma})\\ 
\hline 
\mbox{D0-brane}&\rule{0mm}{6mm}{\cal O}_{x}, x\in \Bn &(0,0,-1) & \delta^{(-1,0)} \\ \hline
\end{array}\nn
\end{eqnarray}

\subsection{Fourier-Mukai transform and $\sl2z$}
 
In this section we will show that a Fourier-Mukai transformation
\cite{RP} on ${\cal B}_{9}=X$ can be identified with an $\sl2z$
transformation by $S=\left({\;0\;\; -1\atop\;1\;\;\;\;0}\right)$ on
the dual 7-brane background ${\bf \widehat{E}_{9}}$.

Let ${\cal F}$ be a complex of sheaves
\footnote{String junctions related by branch cut moves are physically
equivalent.  To a given $(\mbox{r},\Sigma, \mbox{k})$ there
corresponds a family of string junctions related to each other by
branch cut moves.  Since as shown in \cite{sharpe} only the image of
the complex in the K-theory group is physically relevant, this image
is sufficient to construct a member of the corresponding family of
string junctions. These aspects of string junctions and their relation
with derived categories is under investigation.} on $\pi:X\rightarrow
\PP^{1}$ such that $\mbox{ch}({\cal F})= (\mbox{r},\Sigma,
\mbox{k})$. The Fourier-Mukai transform ${\bf S}$ maps this to a
complex of sheaves ${\bf S}({\cal F})$ on $\widehat{X}$, where
$\widehat{\pi}:\widehat{X}\rightarrow \PP^{1}$ is the dual elliptic
fibration.  Let $\widehat{\bf S}$ be the Fourier-Mukai transform which
maps $\widehat{\cal G}$, a complex of sheaves on $\widehat{X}$ with
$\mbox{ch}(\widehat{\cal G})=
(\widehat{\mbox{r}},\widehat{\Sigma},\widehat{\mbox{k}})$, to
$\widehat{\bf S}(\widehat{\cal G})$, a complex of sheaves on $X$.  The
Chern classes of these complexes are given by \cite{RP}
\begin{eqnarray} 
\begin{array}{llllcl}
\mbox{ch}_{0}({\bf S}({\cal F}))\equiv \widehat{\mbox{r}}'=\mbox{d}_{\Sigma},\\ 
\mbox{ch}_{1}({\bf S}({\cal F}))\equiv \widehat{\Sigma}'
=-\mbox{w}(\Sigma)+(\mbox{d}_{\Sigma}-\mbox{r})B
+\{\mbox{k}+\inter{\Sigma}{B}+\frac{1}{2}\mbox{d}_{\Sigma}\}F,\\
\mbox{ch}_{2}({\bf S}({\cal F}))\equiv \widehat{\mbox{k}}'
=-\mbox{d}_{\Sigma}-\inter{\Sigma}{B}+ \frac{1}{2}\mbox{r},
\end{array}
\end{eqnarray} 
and 
\begin{eqnarray} 
\begin{array}{llllcl}
\mbox{ch}_{0}(\widehat{\bf S}(\widehat{\cal G}))\equiv \mbox{r}'=\mbox{d}_{\widehat{\Sigma}},\\ 
\mbox{ch}_{1}(\widehat{\bf S}(\widehat{\cal G}))\equiv \Sigma'
=\mbox{w}^{-1}(\widehat{\Sigma})-(\mbox{d}_{\widehat{\Sigma}}+\widehat{\mbox{r}})B
+\{\widehat{\mbox{k}}-\inter{\widehat{\Sigma}}{B}-\frac{1}{2}\mbox{d}_{\widehat{\Sigma}}\}F,\\
\mbox{ch}_{2}(\widehat{\bf S}(\widehat{\cal G}))\equiv \mbox{k}'
=-\mbox{d}_{\widehat{\Sigma}}-\inter{\widehat{\Sigma}}{B}- \frac{1}{2}\widehat{\mbox{r}},,
\end{array}
\end{eqnarray} 
where $\mbox{w}:H_{2}(X,\bbbz)\rightarrow H_{2}(\widehat{X},\bbbz)$ is
an automorphism induced by isomorphism between $X$ and
$\widehat{X}$. This automorphism is such that $\mbox{w}(B)=B$ and
$\mbox{w}(F)=F$, therefore it corresponds to a Weyl transformation on
the $E_{8}$ root lattice, $\Gamma_{E_{8}}\subset H_{2}(X,\bbbz)$.
 
Thus we see that
\begin{eqnarray}
\pmatrix{\mbox{d}_{\widehat{\Sigma}'}\cr\widehat{\mbox{r}'}} = 
\pmatrix{0&-1\cr 1&~~0}
\pmatrix{\mbox{d}_{\Sigma}\cr\mbox{r} },\hspace{.3in}  
\pmatrix{\mbox{d}_{\Sigma'}\cr\mbox{r}'} = \pmatrix{0&-1\cr 1&~~0}
\pmatrix{\mbox{d}_{\widehat{\Sigma}}\cr\widehat{\mbox{r}} },  
\end{eqnarray}
and
\begin{eqnarray}
\mbox{ch}(\widehat{\bf S}({\bf S}({\cal F})))&=&- \mbox{ch}({\cal F})\\
\langle \,\widehat{\bf S}(\widehat{\cal G}), 
\widehat{\bf S}(\widehat{\cal G})\,\rangle&=&
\langle\, {\bf S}({\cal F}), {\bf S}({\cal F})\,
\rangle =\langle {\cal F}, {\cal F}\rangle .
\end{eqnarray}
If we denote the junction corresponding to ${\cal F}$ by $\mJ_{\cal
F}$,  then\footnote{Strictly speaking the junctions $\mJ_{\cal F}$ and 
 $\mJ_{{\bf S}({\cal F})}$ belong to different 7-brane configurations, 
since under $S$ transformation the 7-brane labels have not changed but the 
$\tau$ parameter of the elliptic curve $E_{*}$ has.}
\begin{eqnarray}
\mJ_{\cal F}&=&\mbox{$\sum_{i=1}^{8}$}
\lambda^{\Sigma}_{i}\omega^{i}+
\mbox{d}_{\Sigma}\omega^{p}+\mbox{r}\omega^{q}+
\inter{\Sigma}{B}\delta^{(0,1)}-
(\mbox{r}+\mbox{k}+\mbox{$\frac{1}{2}$}\mbox{d}_{\Sigma})\delta^{(-1,0)},
\nn\\
\mJ_{{\bf S}({\cal F})}&=&-\mbox{$\sum_{i=1}^{8}$}
\mbox{w}(\lambda^{\Sigma}_{i})\omega^{i}-
\mbox{r}\omega^p+ \mbox{d}_{\Sigma}\omega^{q}+ \{\mbox{r}+\mbox{k}-
\mbox{$\frac{1}{2}$}
\mbox{d}_{\Sigma}\}\delta^{(0,1)}+\inter{\Sigma}{B}\delta^{(-1,0)},  \nn 
\end{eqnarray}
and 
\begin{eqnarray}
\mJ_{\widehat{\cal G}}&=&\mbox{$\sum_{i=1}^{8}$}
\lambda^{\widehat{\Sigma}}_{i}\omega^{i}
+\mbox{d}_{\widehat{\Sigma}}\omega^{p}+
\widehat{\mbox{r}}\omega^{q}+\inter{\widehat{\Sigma}}{B}\delta^{(0,1)}-
(\widehat{\mbox{r}}+\widehat{\mbox{k}}+
\mbox{$\frac{1}{2}$}\mbox{d}_{\widehat{\Sigma}})\delta^{(-1,0)},\\
\mJ_{\widehat{\bf S}(\widehat{\cal G})}&=&\mbox{$\sum_{i=1}^{8}$}
\mbox{w}^{-1}(\lambda^{\widehat{\Sigma}}_{i})\omega^{i}
-\widehat{\mbox{r}}\omega^{p}+ \mbox{d}_{\widehat{\Sigma}}\omega^{q}+
\{\widehat{\mbox{r}}+\widehat{\mbox{k}}+
\mbox{$\frac{1}{2}$}
\mbox{d}_{\widehat{\Sigma}}\}\delta^{(0,1)}+ 
\{\inter{\widehat{\Sigma}}{B}+\widehat{\mbox{r}}\} 
\delta^{(-1,0)}. \nn 
\end{eqnarray}
It is straightforward to show that $\mJ_{{\bf S}({\cal F})}$ and
$\mJ_{\widehat{\bf S}(\widehat{\cal G})}$ are obtained from $\mJ_{\cal
F}$ and $\mJ_{\widehat{\cal G}}$ respectively by a global $\sl2z$
transformation by $S=\left({\;0\;\; -1\atop\;1\;\;\;\;0}\right)$ and
branch cut moves inducing the $E_{8}$ Weyl transformation
$\mbox{w}$. The branch cut moves, however, are different for $
\mJ_{{\bf S}({\cal F})}$ and $\mJ_{\widehat{\bf S}(\widehat{\cal G})}$
and correspond to different Weyl transformations of affine $E_{8}$
\cite{newpaper}. Therefore we see that the Fourier-Mukai transform is
the S-duality transformation of the type IIB 7-brane background ${\bf
\widehat{E}_{9}}$, and the sign ambiguity referred to in
\cite{sharpe,Aspinwall} is required for the proper identification
between junctions and vector bundles.

\section*{Acknowledgements}
We would like to thank Oliver DeWolfe, Robbert Dijkgraaf, David
Morrison, Jun Song, Cumrun Vafa and Barton Zwiebach for valuable
discussions. We are grateful to D. Hern\'andez Ruip\'erez for
providing us with the updated version of their paper. This research was
supported in part by the US Department of Energy under contract
\#DE-FC02-94ER40818.

\end{document}